\documentclass[twocolumn,aps,prd,preprintnumbers,superscriptaddress,nofootinbib,10pt]{revtex4-1}
\usepackage{epsfig}
\usepackage{graphicx}
\usepackage{psfrag}
\usepackage{amsmath,amssymb}
\usepackage{colordvi}
\usepackage{amsfonts}
\usepackage{enumerate}
\usepackage{slashed}
\usepackage{color}
\usepackage{xcolor}

\begin{document}

\title{Single Transverse-Spin Asymmetry and Sivers Function \\ in Large Momentum Effective Theory}

\author{Xiangdong Ji}
\affiliation{Center for Nuclear Femtography, SURA, 1201 New York Ave. NW, Washington, DC 20005, USA}
\affiliation{Maryland Center for Fundamental Physics,
Department of Physics, University of Maryland,
College Park, Maryland 20742, USA}
\author{Yizhuang Liu}
\affiliation{Tsung-Dao Lee Institute, Shanghai Jiao Tong University, Shanghai 200240, China}
\affiliation{Institut f\"{u}r Theoretische Physik, Universit\"{a}t Regensburg, D-93040 Regensburg, Germany}
\author{Andreas Sch\"{a}fer}
\affiliation{Institut f\"{u}r Theoretische Physik, Universit\"{a}t Regensburg, D-93040 Regensburg, Germany}
\author{Feng Yuan}
\affiliation{Nuclear Science Division, Lawrence Berkeley
National Laboratory, Berkeley, CA 94720, USA}

\date{\today}
\vspace{0.5in}
\begin{abstract}
 We apply recent developments in large momentum effective theory (LaMET) to formulate a non-perturbative 
 calculation of the single-transverse spin asymmetry in terms of the quasi transverse-momentum-dependent quark distribution functions from the so-called Sivers mechanism. When the spin asymmetry is defined as the ratio of the quark Sivers function over the spin averaged distribution, it can be directly calculated in terms of the relevant quasi distributions with the soft functions and perturbative matching kernels cancelling out. Apart from the general formula presented, we have verified the result in the small transverse distance limit
 at one-loop order, which reduces to a collinear expansion at twist-three level. 
\end{abstract}

\maketitle

\section{Introduction}
Transverse momentum dependent (TMD) parton distribution functions (PDFs, TMDPDFs) are one of the important ingredients in nucleon tomography and a central focus of hadron physics research in recent years and especially at the future electron-ion collider~\cite{Accardi:2012qut,Boer:2011fh}. TMDPDFs can be experimentally extracted from hard processes in deep inelastic scattering (DIS) and lepton pair production in hadronic collisions~\cite{Collins:2011zzd,Collins:1981uk,Ji:2004wu,Bacchetta:2008xw}. The available 
experimental data and global analysis have generated strong interest in the hadron physics community, see, e.g., recent efforts in Refs.~\cite{Cammarota:2020qcw,Bacchetta:2020gko}.

The first attempt to compute the moments of TMDPDFs from lattice QCD has been made in 
Refs.~\cite{Musch:2010ka,Musch:2011er,Yoon:2017qzo}. Meanwhile, great progress has been made to compute 
$x$-dependent parton physics on the lattice using large momentum effective theory (LaMET)~\cite{Ji:2013dva,Ji:2014gla}, see, some recent reviews on this topic~\cite{Cichy:2018mum,Ji:2020ect}. LaMET is based on the observation that parton physics defined in terms of lightcone correlations can be obtained from  time-independent Euclidean correlations (called quasi-distributions) through well-defined effective field theory (EFT) expansion as well as matching and running. LaMET has been applied to compute various collinear PDFs and distribution amplitudes~\cite{Cichy:2018mum,Ji:2020ect}. In the last few years, an
important new development has been to apply LaMET to describe TMDPDFs and associated soft functions~\cite{Ji:2014hxa,Ji:2018hvs,Ebert:2018gzl,Ebert:2019okf,Ebert:2019tvc,Shanahan:2019zcq,Shanahan:2020zxr,Ebert:2020gxr,Ji:2019sxk,Ji:2019ewn,Vladimirov:2020ofp,Zhang:2020dbb}. In this paper, we study 
single transverse-spin asymmetries in the region where the transverse momentum is on the order of $\Lambda_{\rm QCD}$, focusing on the non-perturbative calculation of the relevant TMDPDF---the quark Sivers function~\cite{Sivers:1989cc}---
in terms of a Euclidean-space quasi distribution.

The spin-dependent, $k_\perp$-even TMDPDFs have been studied in Ref.~\cite{Ebert:2020gxr}, where similar factorization and matching were found as for the unpolarized case. Because the quark Sivers function is a $k_\perp$-odd distribution, it has special features different from those of the $k_\perp$-even ones. 
In particular, in the large $k_\perp$ or small transverse distance limit, the quark Sivers function can be expressed in terms of the collinear twist-three quark-gluon-quark correlation functions in the nucleon, whereas the $k_\perp$-even TMDPDFs depend on the leading twist collinear quark distribution functions. Therefore, the EFT matching calculation in the present case is more involved compared with that in Ref.~\cite{Ebert:2020gxr}.

In this paper, we will focus on computing the quark-Sivers function 
in the leading-order expansion from large-momentum effective theory. 
An extension to the gluon-Sivers function should be possible in a similar manner. The quark-Sivers function describes a nontrivial correlation between the quark's transverse momentum and the nucleon's transverse polarization vector. Therefore, it represents a spin asymmetry in the TMDPDF. The quark Sivers function is non-zero because the gauge link associated with the quark distribution contributes the phase needed to obtain a single spin asymmetry~\cite{Brodsky:2002cx,Collins:2002kn,Ji:2002aa,Belitsky:2002sm}. 

The paper is organized as follows. In Sec.~II, we present the evolution equation of the quasi-Sivers function and its matching to the physical Sivers function. Resummation formulas for the quasi-Sivers function and the matching kernel are also given. A generic argument to demonstrate the matching between the quasi-Sivers function and the light-cone Sivers function will be presented based on the factorization of the hard, collinear and soft gluon radiation contributions for the TMDPDFs. Because the matching coefficient only concerns hard gluon radiation, it does not depend on the spin structure of the nucleon. This is consistent with the observation in Ref.~\cite{Ebert:2020gxr}.
In Sec.~III, we provide detailed derivations of the quasi-Sivers function in LaMET up to one-loop order at large transverse momentum. Our calculations is based on the collinear twist-three quark-gluon-quark correlation functions, and we compute the quasi-Sivers function in terms of the Qiu-Sterman matrix element~\cite{Efremov:1981sh,Efremov:1984ip,Qiu:1991pp,Qiu:1991wg,Qiu:1998ia} (defined below). This can be compared to the light-cone quark-Sivers function calculated in the same framework~\cite{Ji:2006ub,Ji:2006vf,Ji:2006br,Koike:2007dg,Kang:2011mr,Sun:2013hua,Scimemi:2019gge} and the associated matching coefficient can be obtained.  In Sec.~IV, we show the application of the formalism to experimental and theoretical single spin asymmetries. Finally, We summarize our paper in Sec. V.

\section{LaMET Expansion of Sivers Function in Euclidean Quasi TMDPDF}

Let us start with the transverse-spin dependent quasi-TMDPDF for quarks in a proton moving along the $+\hat z$ direction~\cite{Ji:2018hvs,Ji:2019ewn}
\begin{align}\label{eq:quasi_tmd}
& \tilde q(x,k_\perp,S_\perp,\mu,\zeta_z)=\int d^2 b_{\perp} ~\int \frac{d\lambda}{2(2\pi)^3}e^{i\lambda x+i\vec{k}_\perp\cdot\vec{b}_\perp}  \\
&\! \lim_{L \rightarrow \infty} \frac{\langle PS| \bar \psi\big(\frac{\lambda n_z }{2}\!+\!\vec{b}_\perp\big)\Gamma{\cal W}_{z}(\frac{\lambda n_z}{2}\!+\!\vec{b}_\perp;-L)\psi\big(\!-\!\frac{\lambda n_z}{2}\big) |PS\rangle}{\sqrt{Z_E(2L,b_\perp,\mu)}} \ , \nonumber
\end{align}
where $\overline{\rm MS}$ renormalization is implied, $b_\perp=|\vec b_\perp|$, and the staple-shaped gauge-link ${\cal W}_z$ is
\begin{align}\label{eq:staplez}
&{\cal W}_z(\xi;-L)=W^{\dagger}_{z}(\xi; -L)W_{\perp}W_{z}(-\xi^zn_z;-L)  \ ,\\
&W_{z}(\xi;-L)= {\cal P}{\rm exp}\Big[-ig\int_{\xi^z}^{-L} ds\, n_z\cdot A(\vec{\xi}_\perp\!+\!n_z s)\Big] \ .
 \end{align}
 The spin dependence is introduced by the hadron state $|PS\rangle$. $x$ and $k_\perp$ are the longitudinal momentum fraction and the transverse momentum carried by the quark, and $\zeta_z=4x^2P_z^2$ is the rapidity or Collins-Soper scale. The direction vector of the gauge-link $n_z$ is defined as $n_z=(0,0,0,1)$ and all coordinates are 4-vectors, e.g. $\vec b_\perp=(0,b_1,b_2,0)$. In contrast, $L$ is just a number. $\mu$ is the ultra-violet (UV) renormalization scale. A transverse gauge link was included to make the gauge links connected. The spin-1/2 proton has momentum $P^z$ 
and is polarized transversely, with the polarization vector $\vec{S}_\perp$ being perpendicular to its momentum direction. The Dirac matrix $\Gamma$ can be chosen as $\Gamma=\gamma^t$ or $\Gamma=\gamma^z$. As we will show, to leading order in $1/P^z$ the two choices are equivalent. The subtraction factor $Z_E(2L,b_\perp,\mu)$ is the vacuum expectation value of a rectangular Wilson-loop that removes the pinch-pole singularity at large $L$~\cite{Ji:2018hvs,Ji:2019ewn}
\begin{align}\label{eq:Z_E}
Z_E(2L,b_\perp,\mu)=\frac{1}{N_c}{\rm Tr}\langle 0|W_{\perp}{\cal W}_z(\vec{b}_\perp;2L)|0\rangle \ .
\end{align}
As emphasized in~\cite{Ji:2018hvs,Ji:2019ewn}, The self-interactions of gauge links are subtracted using $\sqrt{Z_{E}}$ in order to remove the pinch-pole singularities~\cite{Ji:2019sxk} and to guarantee the existence of the large $L$ limit.

With the above definition, we can express the transverse-spin dependent quasi-TMDPDF in terms
of appropriate Lorentz structures,
\begin{align}\label{eq:sivermomen}
&\tilde q(x,k_\perp,S_\perp,\mu,\zeta_z)\nonumber \\ &=\tilde q(x,k_\perp,\mu,\zeta_z)+\frac{\tilde f_{1T}^{\perp}(x,k_\perp,\mu,\zeta_z)\epsilon^{\beta\alpha}S_{\perp\beta}k_{\perp\alpha}}{M_P} \ ,
\end{align}
where $M_P$ is the proton mass and $\epsilon^{12}=1$ in our convention. In the above equation,
the first term represents the spin-averaged, unpolarized quark distribution and the second term is the quark Sivers function in LaMET. It is also convenient to Fourier transform the $k_\perp$ distribution to get the $b_\perp$-space expression,
\begin{align}
\tilde q(x,b_\perp,S_\perp,\mu,\zeta_z)=\int d^2k_\perp e^{-i\vec{k}_\perp\cdot \vec{b}_\perp} \tilde q(x,k_\perp,S_\perp,\mu,\zeta_z) \ ,
\end{align}
which is convenient for factorization calculations. 
We can similarly express the quasi-TMDPDF in $b_\perp$-space as,
\begin{align}\label{eq:decompo}
&\tilde q(x,b_\perp,S_\perp,\mu,\zeta_z)\nonumber \\ &=\tilde{q}(x,b_\perp,\mu,\zeta_z)+\epsilon^{\alpha\beta}S_\perp^\beta
\tilde{f}_{1T}^{\perp\alpha}(x,b_\perp,\mu,\zeta_z)  \ .
\end{align}
We would like to point out that $\tilde{q}$ is the Fourier transform of the spin-average quark TMDPDF in momentum space, but $\tilde{f}^{\alpha}_{1T}(x,b_\perp,\mu,\zeta_z)$ is not a direct Fourier transform of  $\tilde f_{1T}^{\perp}(x,k_\perp,\mu,\zeta_z)$ due to the presence of $k_{\perp\alpha}$ in Eq.~(\ref{eq:sivermomen}). Our focus in this paper is the large-momentum
factorization of $\tilde{f}_{1T}^{\perp\alpha}(x,b_\perp,\mu,\zeta_z)$.

\subsection{Evolution Equation}
We start with the renormalization property of quasi-TMDPDFs. Similar as for the unpolarized quasi-TMDPDFs, in the numerator of Eq.~(\ref{eq:quasi_tmd}), there are linear divergences associated to the self-energy of the staple-shaped gauge-link ${\cal W}_z$ and logarithmic divergences associated to the quark-link vertices. In addition, there are cusp-UV divergence associated to the junctions between longitudinal and transverse gauge-links at $z=-L$. After subtraction using $\sqrt{Z_E}$ in the denominator, the linear divergences and the cusp-UV divergences all cancel, and one is left with only the logarithmic divergences for quark-link vertices. The associated anomalous dimensions are all equal and are known to be equivalent to the anomalous dimension $\gamma_F$ for the heavy-light quark current~\cite{Falk:1990yz,Ji:1991pr}. It can also be derived as the anomalous dimension of the quark field in $A^z=0$ gauge. Thus, the spin-dependent quasi-TMDPDF, in particular the quasi-Sivers function satisfies the following renormalization group equation 
\begin{align}
\mu^2\frac{d}{d\mu^2}\ln \tilde f_{1T}^{\perp\alpha}(x,b_\perp,\mu,\zeta_z)=\gamma_F \ .
\end{align}
At one-loop level one has $\gamma_F=\frac{3\alpha_sC_F}{4\pi}$ and high-order results can be found in references~\cite{Falk:1990yz,Ji:1991pr}.

We then come to the evolution equation of quasi-TMDPDFs with respect to $\zeta_z$, i.e. the momentum evolution equation~\cite{Ji:2020ect}. Similar to the case of quasi-PDFs or unpolarized quasi-TMDPDFs, at large $P^z$ there are large logarithms of $P^z$ that can be resummed by the corresponding  momentum evolution equation. Using diagramatic methods developed in~\cite{Collins:1981uk,Collins:2011zzd}, it can be shown~\cite{Collins:1981uk,Ji:2014hxa} that $\tilde f_{1T}^{\perp\alpha}$ satisfies the evolution equation
\begin{align}
2\zeta_z\frac{d}{d\zeta_z}\ln \tilde f_{1T}^{\perp\alpha}(x,b_\perp,\mu,\zeta_z)=K(b_\perp,\mu)+G(\zeta_z,\mu)\ ,
\end{align}
where $K$ is the non-perturbative Collins-Soper kernel~\cite{Collins:1981uk} and the $G$ is a perturbative part of the evolution kernel.  At one-loop level, one has~\cite{Collins:1981uk,Ji:2014hxa},
\begin{align}\label{eq:oneloopkg}
K^{(1)}(b_\perp,\mu)&=-\frac{\alpha_sC_F}{\pi}L_b\, ,\\
G^{(1)}(\zeta_z,\mu)&=\frac{\alpha_sC_F}{\pi}\left(1-L_z\right) \, .
\end{align}
Here $L_b=\ln \frac{\mu^2 b_\perp^2}{c_0^2}$ with $c_0=2e^{-\gamma_{E}}$ and $L_z=\ln \frac{\zeta_z}{\mu^2}$. These equations allows the quasi-Sivers function to be re-summed in the form in which $\mu_b=\frac{c_0}{b_\perp}$
\begin{align}
\tilde f_{1T}^{\perp\alpha}(x,b_\perp,\mu,\zeta_z)=\tilde f_{1T}^{\perp\alpha}(x,& b_\perp,\mu=\sqrt{\zeta_z}=\mu_b)\nonumber\\ \times \exp \bigg(\frac{1}{2}\ln\frac{\zeta_z}{\mu_b^2}K(b_\perp,\mu)&+\int_{\mu_b^2}^{\zeta_z} \frac{d\zeta'}{2\zeta'}G(\zeta',\mu)\nonumber \\
&+\int^{\mu}_{\mu_b}\frac{d\mu'^2}{\mu'^2}
\gamma_F(\alpha_s(\mu'))\bigg) 
\end{align}
with $\mu_b=\frac{c_0}{b_\perp}$.
Using the renormalization group equation~\cite{Collins:1981uk,Ji:2019ewn} for $G$ and $K$, $\frac{d\ln G}{d\ln \mu} = -\frac{d\ln K}{d\ln \mu}=2\Gamma_{\rm cusp}$ where $\Gamma_{\rm cusp}$ is the light-like cusp-anomalous dimension,  allows a more refined treatment of resummation for $K$ and $G$.

\subsection{LaMET Expansion for Sivers Function}

Similar to the unpolarized case~\cite{Ji:2019ewn}, the quasi-TMDPDFs can be used to calculate the physical TMDPDFs appearing in the factorizations of experimental cross sections. The LaMET expansion formula requires the off-light-cone reduced soft function $S_r(b_\perp,\mu)$, the definition and more properties of which can be found in~\cite{Ji:2019ewn,Ji:2019sxk,Ji:2020ect}.
In terms of the non-perturbative reduced soft function, the EFT expansion formula for spin-dependent quasi-TMDPDF reads:
\begin{align}\label{eq:factorization}
&f_{1T}^{\perp\alpha}(x,b_\perp,\mu,\zeta)\nonumber \\ &=\frac{ e^{-K(b_\perp,\mu)\ln (\frac{\zeta_z}{\zeta})}} {H\left(\frac{\zeta_z}{\mu^2}\right)}\sqrt{S_r(b_\perp,\mu)} \tilde f_{1T}^{\perp\alpha}(x,b_\perp,\mu,\zeta_z) + ...\ .
\end{align}
where $H$ is the perturbative kernel and the higher-order terms in $1/P^z$ expansoin
have been omitted.

Similar to the unpolarized case~\cite{Ji:2019ewn}, here we provide the sketch of a proof for the matching formula~(\ref{eq:factorization}). We also argue that $H$ is independent of the spin structure, as was recently argued in Ref.~\cite{Ebert:2020gxr}. 

\begin{figure}[t]
\includegraphics[width=0.7\columnwidth]{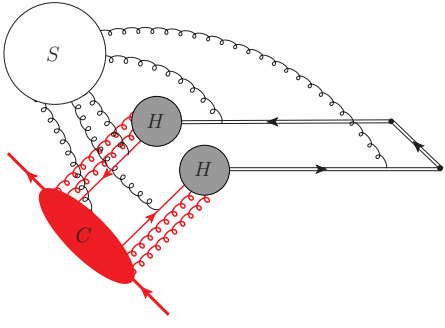}
\caption{The leading regions of the quasi-TMDPDF where $C$ is the collinear subdiagram, $S$ is the soft subdiagram and $H$'s are hard subdiagrams. The two hard cores are not connected with each other (but their open Dirac indices are ontracted), and as a result, the momentum fraction of the quasi-TMDPDF receives only contributions from collinear modes and there is no convolution in the matching formula. }
\label{fig:quasireduce}
\end{figure}

First of all, one can perform a standard leading region analysis~\cite{Collins:2011zzd} for all spin-structures with minor modifications to include the staple-shaped gauge-links of the quasi-TMDPDF as in~\cite{Ji:2019ewn}. The leading region or the reduced diagram for quasi-TMDPDFs is shown in Figure.~\ref{fig:quasireduce}. There are collinear and soft subdiagrams responsible for collinear and soft contributions. The collinear contributions are exactly the same as those for the light-cone TMDPDF defined with light-like gauge-links. The soft radiations between the fast moving color charges and the staple shaped gauge-links can be factorized by the off-light-cone soft function. 

In addition to the collinear and soft subdiagrams, there are two hard subdiagrams around the vertices at $0$ and $\vec{b}_\perp$. The natural hard scale $\zeta_z$ for the hard diagram is formed by a Lorentz invariant combination of the parton momenta entering the hard subdiagram and the direction vector $n_z$ for the staple-shaped gauge links. At large $P^z$, small $k_\perp$ or large $b_\perp$, the hard contributions are confined within the vicinities of the quark-link vertices around $0$ and $b_\perp$, since any hard momenta flowing between $0$ and $b_\perp$ will cause additional power suppressions in $\frac{1}{P^z}$. In another words, there are two disconnected hard subdiagrams, one containing $0$ and another one containing $\vec{b}_\perp$. Therefore, the momentum fractions carried by the quasi-TMDPDF and the physical TMDPDF are the same and the matching formula contains no convolution. 

Given the leading region of the quasi-TMDPDF, one can apply the standard Ward-identity argument of Ref.~\cite{Collins:2011zzd} to factorize the quasi-TMDPDF and obtain Eq.~(\ref{eq:factorization}). The reduced soft function, which is actually the inverse of the rapidity independent part of the off-light-cone soft function~\cite{Ji:2019sxk,Ji:2019ewn}, appears to compensate the differences of soft contributions for quasi-TMDPDFs and physical TMDPDFs. The exponential of the Collins-Soper kernel can be explained by the emergence of large logarithms for quasi-TMDPDFs in the form $K(b_\perp,\mu)\ln \frac{\zeta_z}{\mu^2}$ generated by momentum evolution. To match to the physical TMDPDF at rapidity scale $\zeta$, a factor $e^{-K(b_\perp,\mu)\ln (\frac{\zeta_z}{\zeta})}$ is therefore needed. Finally, the mismatch between $\tilde f$ and $f$ due to the hard contributions is captured by the hard kernel $H$ that depends on the hard scale $\zeta_z$ and the renormalization scale $\mu$. The above arguments are similar for the unpolarized case~\cite{Ji:2019ewn}.  

\begin{figure}[t]
\includegraphics[width=0.6\columnwidth]{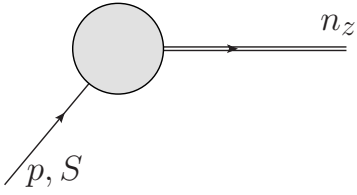}
\caption{The form factor shown here is sufficient for calculating the matching kernel, which 
contains an incoming quark with momentum $p=xP$, spin $S$ and an outgoing gauge-link in $n_z$ direction. }
\label{fig:formfactor}
\end{figure}

Here we argue that the hard kernel $H$ is independent of the spin structure. As we already emphasized, the hard-cores around $0$ and $\vec{b}_\perp$ are disconnected. Any momentum that is allowed to flow between the vertices and sees the transverse separation is either soft or collinear. The hard momenta have essentially no effects on the other vertex. Therefore, in order to obtain the matching kernel, it is sufficient to consider only ``half'' of the quark quasi-TMDPDF, which one might want to call an ``amputated'' form factor containing only an incoming light-quark with momentum $p=xP$ and an ``outing going'' gauge-link along $n_z$ direction. This form factor is shown in Fig.~\ref{fig:formfactor}. For this form factor, the generic Lorentz structure can always be written as
\begin{align}
\Gamma\left(A +B \gamma \cdot n_z \gamma \cdot p +C\gamma \cdot p \gamma \cdot n_z \right) u(p,S)\ ,
\end{align}
where $\Gamma$ is a generic Dirac matrix at the quark-link vertex, unrelated to that in Eq.(1), $u(p,S)$ is the Dirac spinor for the incoming quark and $A$,$B$,$C$ are scalar functions of $p^2$, $n_z^2$ and $n_z \cdot p$ . Using the anti-commutation relation of Dirac matrices and the equation of motion $\gamma \cdot p u(p,S)=0$, the above equation can be rewritten as
\begin{align}
\Gamma\left(A +2Cn_z\cdot p\right) u(p,S) \ ,
\end{align}
which depends only on a universal scalar function $A +2Cn_z\cdot p$, independent of the spin $S$ and the Dirac matrix $\Gamma$. As a result, the matching kernel only depends on these scalar functions but not the spin $S$ and the Dirac matrix $\Gamma$.


The above general results can be verified at one-loop order when $b_\perp$ is small
and a perturbative QCD calculation is valid. The one-loop 
calculation is more complicated compared with that in~\cite{Ebert:2020gxr}, because
it involves twist-three collinear factorization. 

First of all, let us recall that the standard TMDPDF factorization at small $b_\perp$ follows the procedure in Refs.~\cite{Collins:2011zzd,Catani:2000vq,Catani:2013tia,Prokudin:2015ysa}. For the quark Sivers function, we have~\cite{Sun:2013hua,Scimemi:2019gge},
\begin{align}\label{eq:oneloopsiver}
&{f}_{1T}^{\perp\alpha}(x,b_\perp,\mu,\zeta)=\frac{ib_\perp^\alpha}{2}T_F(x,x)\nonumber \\ &+\frac{ib_\perp^\alpha}{2}\frac{\alpha_s}{2\pi}\left\{\left(-\frac{1}{\epsilon}
-L_b\right){\cal P}_{qg/qg}^T\otimes T_F(x,x)\right.\nonumber\\
&\left.+\int\frac{dx_q}{x_q}T_F(x_q,x_q)\left[-\frac{1}{2N_c}(1-\xi_x)+\delta(1-\xi_x)C_Fs^{(1)}\right]\right\}\ ,
\end{align}
where $\xi_x=\frac{x}{x_q}$, $T_F(x,x)$ is the twist-3 quark-gluon-quark correlation function (the Qiu-Sterman matrix element) defined below and ${\cal P}_{qg/qg}^T$ is the associated splitting kernel. For the part involved in the calculations of Sec.~III, we have~\cite{Braun:2009mi,Schafer:2012ra,Kang:2012em,Sun:2013hua,Scimemi:2019gge}~\footnote{Here and in the following calculations, we only keep the so-called soft gluon pole and hard gluon pole contributions in the twist-three formalism~\cite{Ji:2006ub,Ji:2006vf,Ji:2006br}. A complete kernel including soft-fermion pole contributions and other twist-3 matrix elements can be found in Refs.~\cite{Braun:2009mi,Scimemi:2019gge}.},
\begin{eqnarray}
&&{\cal P}_{qg/qg}^T\otimes T_F(x,x)\nonumber\\
&&~=\int\frac{dx_q}{x_q} \left\{T_F(x_q,x_q)\left[C_F\left(\frac{1+\xi_x^2}{1-\xi_x}\right)_+-C_A\delta(1-\xi_x)\right]\right.\nonumber\\
&&~~\left.
+\frac{C_A}{2}\left(T_F(x_q,x)\frac{1+\xi_x}{1-\xi_x}-T_F(x_q,x_q)\frac{1+\xi_x^2}{1-\xi_x}\right)\right\}  \ .\label{ptt}
\end{eqnarray}
The contribution $s^{(1)}$ reads
\begin{align}
s^{(1)}=-\frac{\pi^2}{12}+\frac{3}{2}L_b-\frac{1}{2}L_b^2-L_{\zeta}L_b \ ,
\end{align}
where $L_{\zeta}=\ln \frac{\zeta}{\mu^2}$. Our definition of the physical TMD-PDF follows the standard one in  Ref.~\cite{Echevarria:2012js,Collins:2012uy}, although the numerical factors $\frac{\pi^2}{12}$ depends on the renormalization schemes due the presence of double $\frac{1}{\epsilon^2}$ poles, see section VI of Ref.~\cite{Collins:2012uy} for a discussion.  Our results are in the standard $\overline {MS}$ scheme.

In the next section, we show that the quasi-Sivers function has a similar factorizaiton
at small $b_\perp$,
\begin{align}\label{eq:oneloopquasisiver}
&{\tilde f}_{1T}^{\perp\alpha}(x,b_\perp,\mu,\zeta_z)=\frac{ib_\perp^\alpha}{2}T_F(x,x)\nonumber \\ &+\frac{ib_\perp^\alpha}{2}\frac{\alpha_s}{2\pi}\left\{\left(-\frac{1}{\epsilon}
-L_b\right){\cal P}_{qg/qg}^T\otimes T_F(x,x)\right.\nonumber\\
&\left.+\int\frac{dx_q}{x_q}T_F(x_q,x_q)\left[-\frac{1}{2N_c}(1-\xi_x)+\delta(1-\xi_x)C_F\tilde s^{(1)}\right]\right\}\ .
\end{align}
where $\tilde s^{(1)}$ reads
\begin{align}
\tilde s^{(1)}=&-2+\frac{5}{2}L_b-\frac{1}{2}L_b^2-L_zL_b-\frac{1}{2}L_z^2+L_z \ .
\end{align}
Furthermore, the one-loop reduced soft function reads~\cite{Ji:2020ect}:
\begin{align}
S_r(b_\perp,\mu)=1-\frac{\alpha_sC_F}{\pi}L_b \ .
\end{align}
Combining all of the above and comparing it to the LaMET expansion in Eq. (\ref{eq:factorization}), one obtains the one-loop matching kernel
\begin{align}
H\left(\frac{\zeta_z}{\mu^2}\right)=1+\frac{\alpha_sC_F}{2\pi}\bigg(-2+\frac{\pi^2}{12}-\frac{1}{2}L_z^2+L_z\bigg) \ ,
\end{align}
which is exactly the answer we expected: The matching kernel is independent of spin-structure and is equal to that of the unpolarized case for $\Gamma=\gamma^z$ and $\Gamma=\gamma^t$.

It can be shown~\cite{Ji:2019ewn} that the matching kernel satisfies the renormalization group equation
\begin{align}
\mu\frac{d}{d\mu} \ln H\left(\frac{\zeta_z}{\mu^2}\right)=\Gamma_{\rm cusp} \ln \frac{\zeta_z}{\mu^2} + \gamma_C
\end{align}
where $\gamma_C$ can be found in Ref.~\cite{Ji:2019ewn}.
The general solution to the above equation reads
\begin{align}\label{eq:C_renormalization}
&H\left(\alpha_s(\mu),\frac{\zeta_z}{\mu^2}\right)=H\left(\alpha_s(\sqrt{\zeta_z}), 1\right)\\
&\times\exp\left\{\int_{\sqrt{\zeta_z}}^{\mu}\frac{d\mu'}{\mu'}\left[\Gamma_{\rm cusp}(\alpha_s(\mu')) \ln \frac{\zeta_z}{\mu^{'^2}}+\gamma_C\big(\alpha_s(\mu')\big)\right]\right\}\nonumber\,.
\end{align}
This equation allows the determination of the large logarithms for $H$ to all orders in perturbation theory, up to unknown constants related to the initial condition $H(\alpha_s,1)$.

\section{One-loop Calculation for Spin-Dependent quasi-TMPPDFs}

In this section we calculate the quasi-Sivers function at one-loop level. The idea and procedure is the same as for previous examples in the LaMET formalism~\cite{Ji:2014hxa,Ji:2018hvs,Ebert:2018gzl,Ebert:2019okf,Ebert:2019tvc,Ebert:2020gxr,Ji:2019sxk,Ji:2019ewn,Vladimirov:2020ofp}. An important difference is that we will not be able to formulate it in terms of a single quark target. Instead, we need to use the collinear twist-three quark-gluon-quark correlation description and compute the quark quasi-TMDPDF and Sivers asymmetry in these collinear 
quark distributions at small $b_\perp\ll 1/\Lambda_{\rm QCD}$.

For the quasi-TMDPDFs, we follow the definition of Eq.~(\ref{eq:quasi_tmd})~\cite{Ji:2019ewn}, where a rectangular Wilson-loop was adopted to remove the pinch-pole singularities. To match to the physical TMDPDFs
at leading order in $1/P^z$, one needs the reduced soft function at small $b_\perp$, which can also be extracted from lattice simulations at any $b_\perp$~\cite{Zhang:2020dbb}. 

The perturbative quasi-TMDPDFs at small $b_\perp\ll 1/\Lambda_{\rm QCD}$ can be expressed in terms of the collinear parton distribution and/or the twist-three quark-gluon-quark correlation functions. For the unpolarized quark distribution, the previous results of \cite{Ji:2018hvs} can be expressed as,
\begin{align}
&\tilde{q}(x,b_\perp,\mu,\zeta_z)\nonumber \\ &=f_q(x_,\mu)+\frac{\alpha_s}{2\pi}\left\{\left(-\frac{1}{\epsilon}
-L_b\right){\cal P}_{q/q}\otimes f_q(x)\right.\nonumber\\
&\left.+C_F\int\frac{dx_q}{x_q}f_q(x_q)\left[(1-\xi_x)+
\delta(1-\xi_x)\tilde s^{(1)}\right]\right\}\ ,
 \label{oneloopun}
\end{align}
for the leading order plus next-to-leading order result in ${\vec b}_\perp$-space, where $\xi_x=x/x_q$, $\mu$ is the renormalization scale in the $\overline{\rm MS}$ scheme, ${\cal P}_{q/q}(\xi_x)=C_F\left(\frac{1+\xi_x^2}{1-\xi_x}\right)_+$ is the usual splitting kernel for the quark, and $f_q(x)$ represents the light-cone integrated quark distribution function. The one-loop coefficient in the subtraction scheme of Ref.~\cite{Ji:2018hvs} reads
\begin{align}
\tilde s^{(1)}=&\frac{3}{2}\ln\frac{b_\perp^2\mu^2}{c_0^2}+\ln\frac{\zeta_zL^2}{4c_0^2}
-\frac{1}{2}\left(\ln\frac{\zeta_zb_\perp^2}{c_0^2}\right)^2\nonumber \\ &+2{\cal K}(\xi_b)-{\cal K}(2\xi_b) \ ,
\end{align}
where the Collins-Soper scale $\zeta_z=4x_q^2P_z^2$ and the function ${\cal K}$ will be defined later on. At large $L$, all the $L$ dependencies cancel and we have:
\begin{align}
\tilde s^{(1)}=-2+\frac{5}{2}L_b-\frac{1}{2}L_b^2-L_zL_b-\frac{1}{2}L_z^2+L_z \ ,\label{h1}
\end{align}
where $L_b$ and $L_z$ are defined after Eq.~(\ref{eq:oneloopkg}).

After the renormalization of the integrated quark distribution $f_q(x)$ at $\mu=\mu_b$, we can write the quasi-TMD unpolarized quark distribution as
\begin{align}
\tilde{q}(x,b_\perp,\mu_b,\zeta_z)&\nonumber \\ =\int\frac{dx_q}{x_q}f_q(x_q,\mu_b)&\bigg(\delta(1-\xi_x)+\frac{\alpha_sC_F}{2\pi}\bigg[(1-\xi_x) \nonumber \\
&+\delta(1-\xi_x)\tilde s^{(1)}\bigg] \bigg)\ .
\end{align}
The goal of the following derivations in this section is to apply the collinear twist-three formalism and calculate the quasi-Sivers function in LaMET at leading order and next-to-leading order.

\subsection{Phase Contribution from the Gauge Link}
In Eq. (\ref{eq:quasi_tmd}), the quasi-TMDPDF contains a gauge link with a finite length, implying that the eikonal gauge link propagators will be modified. From previous works~\cite{Brodsky:2002cx,Collins:2002kn,Ji:2002aa,Belitsky:2002sm}, we know that the gauge link propagators contribute to the crucial phase which is necessary to generate a non-zero Sivers function. Therefore, we need to check that the finite length gauge link can still do so.

Because of the finite length of the gauge links, the eikonal propagator
in these diagrams will be modified according to
\begin{equation}
(-ig)\frac{i n^\mu }{n\cdot k\pm i\epsilon}\Longrightarrow
(-ig)\frac{i n^\mu}{n\cdot k}\left(1-e^{\pm in\cdot k L}\right) \ ,\label{eikonal}
\end{equation}
where $n^\mu$ represents the gauge link direction. In the current case $n^\mu=n_z^\mu$.
In perturbation calculations, we will make use of the large
length limit $|LP_z|\gg 1$. By doing so, many previous results can
be applied to our calculations.
For example, in the large $L$ limit, we have the
following identity:
\begin{equation}
    \lim_{L\to \infty}\frac{1}{n\cdot k}e^{\pm i Ln\cdot k}=\pm i\pi \delta(n\cdot k) \ ,
\end{equation}
which will contribute to the phases needed for a non-zero quark Sivers function.

In the following calculations, we will take two limits whenever this is possible: The large $L$ limit and the large $P_z$ limit. In certain diagrams, we have to use finite $L$ and $P_z$ to regulate, for example, the pinch-pole singularity and/or the end-point singularity~\cite{Ji:2018hvs}. We will emphasize these important points when we carry out the detailed calculations.

\subsection{Leading Order}

We carry out the derivations in the twist-three collinear framework, where the quark Sivers function depends on the so-called twist-three quark-gluon-quark correlation function, aka, the Qiu-Sterman matrix elements~\cite{Efremov:1981sh,Efremov:1984ip,Qiu:1991pp,Qiu:1991wg,Qiu:1998ia}. It is defined as follows,
\begin{align}
&T_F(x_2,x_2') \equiv  \int\frac{d\zeta^-d\eta^-}{4\pi} e^{i(x_2
P^+\eta^-+(x_2'-x_2)P_B^+\zeta^-)}\epsilon_\perp^{\beta\alpha}S_{\perp\beta}
\nonumber \\
&\times   \,
\left\langle PS|\overline\psi(0){\cal L}(0,\zeta^-)\gamma^+
g{F_\alpha}^+ (\zeta^-) {\cal L}(\zeta^-,\eta^-)
\psi(\eta^-)|PS\right\rangle  \ ,  \label{TF}
\end{align}
where $F^{\mu\nu}$ represent the gluon field strength tensor. From the leading order derivation~\cite{Boer:2003cm}, we have,
\begin{equation}
    \frac{1}{M_P}\int d^2k_{\perp}\, k^2_\perp\, f_{1T}^{\perp(\rm SIDIS)}(x,k_\perp) = - T_F(x,x) \ ,\label{moment}
\end{equation}
where $f_{1T}^{\perp({\rm SIDIS})}$ represents the quark Sivers function for a SIDIS process with gauge link going to $+\infty$, corresponding to our choice of $-L$ in Eq.~(\ref{eq:quasi_tmd}).

\begin{figure}[tbp]
\begin{center}
\includegraphics[width=7cm]{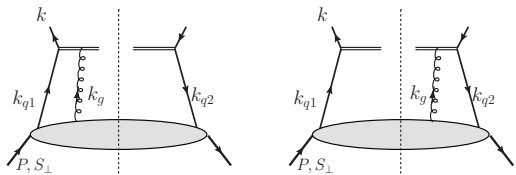}
\end{center}
\caption[*]{Leading order diagrams for quasi-Sivers function.}
\label{leadingorder}
\end{figure}

The method for calculating the single transverse-spin asymmetry in the twist-three formalism has been well developed~\cite{Qiu:1991pp,Qiu:1991wg,Qiu:1998ia,Ji:2006ub,Ji:2006vf,Ji:2006br,Kouvaris:2006zy,Eguchi:2006qz,Eguchi:2006mc,Koike:2006qv,Koike:2007rq,Koike:2007dg,Braun:2009mi,Kang:2008ey,Vogelsang:2009pj,Zhou:2008mz,Schafer:2012ra,Kang:2011mr,Sun:2013hua,Scimemi:2019gge}. There are different approaches to derive the final result, in the following, we follow the collinear $k_{g\perp}$-expansion method~\cite{Qiu:1991pp,Qiu:1991wg,Qiu:1998ia,Ji:2006ub,Ji:2006vf,Ji:2006br,Kouvaris:2006zy}. In this approach, the additional gluon from the polarized hadron is associated with a gauge potential $A^+$, assuming that the polarized nucleon is moving along the $+\hat z$ direction. Thus, the gluon will carry longitudinal polarization and its momentum is parameterized as $x_gP+k_{g\perp}$, where $x_g$ is the momentum fraction with respect to the polarized proton and $k_{g\perp}$ is the transverse momentum. The contribution to the single-transverse-spin asymmetry arises from terms linear in $k_{g\perp}$ in the expansion of the partonic amplitudes. When combined with $A^+$, these linear terms will yield $\partial^\perp A^+$, a part of the gauge field strength tensor $F^{\perp +}$ in Eq.~(\ref{TF}). 
As shown in Fig.~\ref{leadingorder}, we have $\vec{k}_{g\perp}=\vec{k}_{q2\perp}-\vec{k}_{q1\perp}$. Therefore, the $k_{g\perp}$ expansion of the scattering amplitudes can be expresssed in terms of the transverse momenta $k_{q1\perp}$ and $k_{q2\perp}$. The associated quark momenta are parameterized as,
\begin{equation}
k_{q1}=x_{q1}P+k_{q1\perp},~~~k_{q2}=x_{q2}P+k_{q2\perp} \ . \label{e55}
\end{equation}

We compute the quasi-Sivers function defined in Eq.~(\ref{eq:quasi_tmd}) with the Gamma matrix $\Gamma=\gamma^t$ or $\gamma^z$. The results are the same in the leading power of $1/P_z$. The leading order diagrams of Fig.~\ref{leadingorder} can be calculated following the above general procedure. The method is similar to that for the standard quark Sivers function calculation in Ref.~\cite{Kang:2011mr,Sun:2013hua}. In particular, the phase comes from the gauge link propagator,
\begin{eqnarray}
\lim_{L\to \infty}\frac{1}{n_z\cdot k_g}e^{\pm iLn_z\cdot k_g}=\pm i\pi\frac{1}{n_z\cdot P}\delta(x_g) \ ,
\end{eqnarray}
which determines the kinematics for the twist-three Qiu-Sterman matrix element at $T_F(x,x)$. The plus/minus signs correspond to the left and right diagrams where the gluon attaches to the left and right sides of the cut-line, respectively. To calculate the Sivers function in $b_\perp$-space, we need to perform a Fourier transformation with respect to the probing quark transverse momentum $k_\perp$ in Fig.~\ref{leadingorder}. Because of momentum conservation, at leading order, $k_\perp=k_{q2\perp}$ for the left diagram and $k_\perp=k_{q1\perp}$ for the right diagram. As shown above, these two diagrams contribute with opposite sign to the Sivers function. Therefore, the total contribution is proportional to:
\begin{equation}
    \left(e^{i\vec{k}_{q2\perp}\cdot \vec{b}_\perp}-e^{i\vec{k}_{q1\perp}\cdot \vec{b}_\perp}\right)
    \to ib_\perp^\alpha(k_{q2\perp}^\alpha-k_{q1\perp}^\alpha) =ib_\perp^\alpha k_{g\perp}^\alpha\ ,
\end{equation}
in the collinear expansion. As a result, the leading order result for the quark Sivers function in LaMET reads
\begin{eqnarray}
\tilde{f}_{1T}^{\perp\alpha(0)}(x,b_\perp,\mu,\zeta_z)=\frac{ib_\perp^\alpha}{2}T_F(x,x) \ .
\end{eqnarray}
Here, the normalization is consistent with Eq.~(\ref{moment}).

\subsection{One-loop Order from Cut Diagrams}

It has been shown that the quark TMDPDFs in LaMET can be evaluated by the cut diagram approximation~\cite{Ji:2014hxa,Ji:2018hvs}. In particular, if we focus on the kinematic region $0<x<1$, the cut diagram approximation leads to the same results as the complete calculation. In the following, we will apply this approximation to simplify the derivation of the quark Sivers function in LaMET. 

\begin{figure}[tbp]
\begin{center}
\includegraphics[width=4cm]{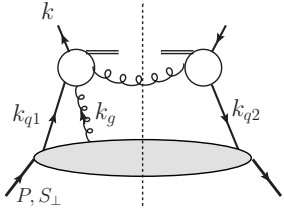}
\end{center}
\caption[*]{Cut diagram approximation to calculate the Siver function in LaMET. A mirror diagram similar to that in Fig.~\ref{leadingorder} should be included as well. The longitudinal gluon from the polarized nucleon can attach to any lines associated with the blob.}
\label{general}
\end{figure}

In Fig.~\ref{general}, we show the generic diagrams to calculate the quark Sivers function in LaMET. The lower part represents the quark-gluon-quark correlation from the polarized nucleon. We follow the strategy of Ref.~\cite{Ji:2006vf} to evaluate these diagrams. The radiated gluon carries transverse momentum $k_{1\perp}$ equal in size but opposite to $k_\perp$. Similar as for the leading diagrams, we need to generate a phase from the gauge link propagators in these diagrams. This corresponds to the pole contributions to the single spin asymmetries in the twist-three formalism~\cite{Ji:2006ub,Ji:2006vf,Ji:2006br}. In the following calculations, we focus on the so-called soft-gluon pole and hard-gluon pole contributions. They are characterized by the longitudinal momentum fraction carried by the gluon attached to the hard partonic part from the polarized nucleon: $x_g=0$ corresponds to the soft-gluon-pole contribution, while $x_g\neq 0$ corresponds to the hard-gluon-pole contribution. It is straightforward to extend this treatment to other contributions such as the soft-fermion pole contribution, and those associated with the twist-three function $\tilde{G}_F$~\cite{Koike:2007dg}.

\begin{figure}[tbp]
\begin{center}
\includegraphics[width=8cm]{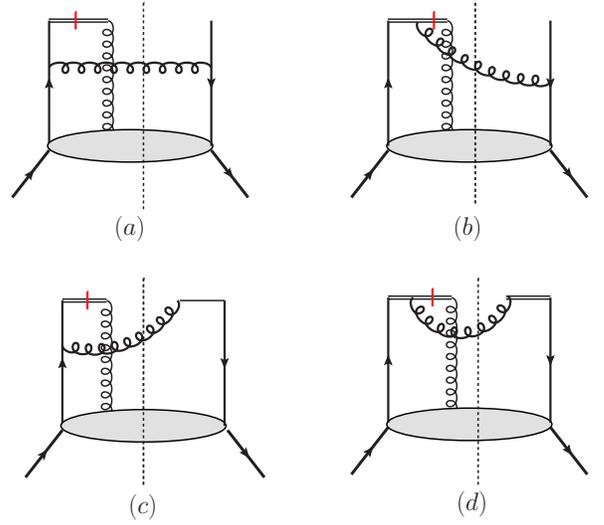}
\end{center}
\caption[*]{Soft-gluonic pole contribution at one-loop order for the real gluon radiation.}
\label{softpole}
\end{figure}

We emphasize again that the quasi-Sivers function defined in Eq.~(\ref{eq:quasi_tmd}) can be computed with $\Gamma=\gamma^t$ or $\gamma^z$ and the results are the same in the TMD limit. The soft gluon pole diagrams are shown in Fig.~\ref{softpole}. The pole contribution is the same as that for the leading order diagrams, i.e., $\delta(n_z\cdot k_g)=\frac{1}{n_z\cdot P}\delta(x_g)$. An important step to obtain the final result is to perform the collinear expansion for the incoming parton momenta. Therefore, we will keep the transverse momenta for $k_{q1}$, $k_{q2}$, and $k_g$. Because of momentum conservation, we have $k_{g\perp}=k_{q2\perp}-k_{q1\perp}$. Therefore, there will be two independent transverse momenta in the expansion. One of the collinear expansion contributions comes from the on-shell condition of the radiated gluon in the cut-diagram approximation. This leads to the so-called derivative terms, which can be easily evaluated~\cite{Ji:2006vf}. The final result can be written as
\begin{align}
&\tilde f_{1T}^\perp(x,k_\perp,\mu,\zeta_z)|_{\partial}=-\frac{M_p}{(k_\perp^2)^2}\frac{\alpha_s}{2\pi^2}\frac{1}{2N_c}\nonumber \\ &\times \int\frac{dx_q}{x_q}\left(x_q\frac{\partial}{\partial x_q}T_F(x_q,x_q)\right)\left(1+\xi_x^2+(1-\xi_x)^2\frac{D-2}{2}\right) \ ,
\end{align}
where $\xi_x=x/x_q$ and $D$ represents the dimension for the transverse plane. In the following we will also use $\epsilon=(2-D)/2$. We have also applied the following relation between the momentum fractions along the $\hat z$ direction and those along the light-cone plus direction,
\begin{eqnarray}
(1-\xi)=(1-\xi_x)\frac{1+\sqrt{1+r_\perp^2}}{2} \ ,
\end{eqnarray}
where $r_\perp=|k_\perp|/(x_q(1-\xi_x)P_z)$. In the TMD limit away from the end-point of $\xi_x=1$, we will have $(1-\xi)\to (1-\xi_x)$. At the end-point, we will have to keep the full expression in order to derive the complete result. However, for the above derivative terms, we can simply substitute $(1-\xi)\to (1-\xi_x)$. We further notice that the derivative terms can be transformed into non-derivative terms by performing an integral by part,
\begin{align}
&\tilde f_{1T}^\perp(x,k_\perp,\mu,\zeta_z)|_{\partial}=-\frac{M_p}{(k_\perp^2)^2}\frac{\alpha_s}{2\pi^2}\frac{1}{2N_c}\int\frac{dx_q}{x_q}T_F(x_q,x_q)\nonumber \\ &\times\left[2\xi_x^2+2\epsilon\xi_x(1-\xi_x)+2\delta(1-\xi_x)\right] \ .
\end{align}
The last term in the square brackets comes from the boundary.

Now, we turn to the non-derivative terms. Fig.~\ref{softpole}(a) is easy to derive because it does not have end-point singularity, and we find that
\begin{align}
&\tilde f_{1T}^\perp(x,k_\perp,\mu,\zeta_z)|_{\rm fig.\ref{softpole}(a)}^{\rm ND}\nonumber \\ &=\frac{M_p}{(k_\perp^2)^2}\frac{\alpha_s}{2\pi^2}\frac{1}{2N_c}\int\frac{dx_q}{x_q}T_F(x_q,x_q)\nonumber\\
&\times \frac{(1-\xi)(1-\epsilon)}{(1-\xi_x)\sqrt{1+r_\perp^2}}\left[(1-2\xi)(1-\xi)+\frac{k_\perp^2}{P_z^2}\right] \ .
\end{align}
Taking the TMD limit, and adding the corresponding term from the non-derivative contribution, we obtain the final result for Fig.~\ref{softpole}(a)
\begin{align}
&\tilde f_{1T}^\perp(x,k_\perp,\mu,\zeta_z)|_{\rm fig.\ref{softpole}(a)}\nonumber \\ &=\frac{M_p}{(k_\perp^2)^2}\frac{\alpha_s}{2\pi^2}\frac{1}{2N_c}\int\frac{dx_q}{x_q}T_F(x_q,x_q) (1-\xi_x)(1-\epsilon) \ .
\end{align}
On the other hand, the diagrams $(b,c)$ of Fig.~\ref{softpole} contribute to the end-point singularities. The result can be written as
\begin{align}
&\tilde f_{1T}^\perp(x,k_\perp,\mu,\zeta_z)|_{\rm fig.\ref{softpole}(b,c)}^{\rm ND}\nonumber \\ &=\frac{M_p}{(k_\perp^2)^2}\frac{\alpha_s}{2\pi^2}\frac{1}{2N_c}\int\frac{dx_q}{x_q}T_F(x_q,x_q)\nonumber\\
&\times\frac{2\xi(1-\xi)^2}{(1-\xi_x)^3\sqrt{1+r_\perp^2}} \left[2-\xi+(1-\xi)r_\perp^2\right] \ .
\end{align}
Clearly, the last term in the bracket is power suppressed in the TMD limit. Furthermore, we can rewrite $[2-\xi]$ as two terms as $1+(1-\xi)$. The first term will have an end-point singularity, whereas the second term is regular. It is interesting to find that this regular term cancels the corresponding term from the derivative contribution derived above. Therefore, there are only end-point contributions from Fig.~\ref{softpole}(b,c),
\begin{align}
&\tilde f_{1T}^\perp(x,k_\perp,\mu,\zeta_z)|_{\rm fig.\ref{softpole}(b,c)}\nonumber \\&=\frac{M_p}{(k_\perp^2)^2}\frac{\alpha_s}{2\pi^2}\frac{1}{2N_c}\int\frac{dx_q}{x_q}T_F(x_q,x_q)\frac{2\xi(1-\xi)^2}{(1-\xi_x)^3\sqrt{1+r_\perp^2}}\ .
\end{align}
We further notice that $(1-\xi)^2$ can be simplified as $(1-\xi)^2= (1-\xi_x)^2(1+\sqrt{1+r_\perp^2})^2/4\approx (1-\xi_x)^2(1+\sqrt{1+r_\perp^2})/2$ in the TMD limit. With that, we obtain
\begin{align}
&\tilde f_{1T}^\perp(x,k_\perp,\mu,\zeta_z)|_{\rm fig.\ref{softpole}(b,c)} \nonumber\\ &=\frac{M_p}{(k_\perp^2)^2}\frac{\alpha_s}{2\pi^2}\frac{1}{2N_c}\int\frac{dx_q}{x_q}T_F(x_q,x_q)\frac{2\xi_x}{1-\xi_x}\frac{1+\sqrt{1+r_\perp^2}}{2\sqrt{1+r_\perp^2}}\nonumber\\
&=\frac{M_p}{(k_\perp^2)^2}\frac{\alpha_s}{2\pi^2}\frac{1}{2N_c}\int\frac{dx_q}{x_q}T_F(x_q,x_q)\nonumber \\ &\times \left[\frac{2\xi_x}{(1-\xi_x)_+}+\delta(1-\xi_x)\ln\frac{\zeta_z}{k_\perp^2}\right]\ ,
\end{align}
where $\zeta_z=4x^2P_z^2$. The last equation follows from a similar derivation for the unpolarized TMD quark calculation in Refs.~\cite{Ji:2018hvs} in the TMD limit.

Fig.~\ref{softpole}(d) is a little more involved, because it has the so-called pinch pole singularity if we take the limit $L\to \infty$ first,
\begin{align}
&\tilde f_{1T}^\perp(x,k_\perp,\mu,\zeta_z)|_{\rm fig.\ref{softpole}(d)}^{L\to \infty}=\nonumber \\ &\frac{M_p}{k_\perp^2}\frac{\alpha_s}{2\pi^2}\frac{1}{2N_c}\int\frac{dx_q}{x_q}T_F(x_q,x_q)\frac{1}{\sqrt{1+r_\perp^2}}\frac{1}{k_{1z}+i\epsilon}\frac{1}{k_{1z}-i\epsilon}\ ,
\end{align}
where $k_{1z}=x_q(1-\xi_x)P_z$ represents the longitudinal momentum carried by the radiated gluon crossing the cut line. Because of the pinch-pole singularity, the above contribution is not well defined around $x=x_q$ ($\xi_x=1$). A finite length of the gauge link will help to regulate the pinch pole singularity as shown for the unpolarized case. In addition, similar to the unpolarized case, the contribution is power suppressed when $x\neq x_q$. Therefore, it will only contribute to a Delta function at $\xi_x=1$. Following the same strategy as in Ref.~\cite{Ji:2018hvs}, we perform the Fourier transformation with respect to $k_\perp$ and carry out the $k_{1z}$ integral to derive the $b_\perp$-space expression. 
Schematically, the quark Sivers function in $b_\perp$-space can be derived as follows,
\begin{align}
&\tilde{f}_{1T}^\perp(x,b_\perp,\mu,\zeta_z)|_{\rm fig.\ref{softpole}(d)}\nonumber \\ & =\int \frac{d^2k_\perp}{(2\pi)^2} e^{i\vec{k}_\perp\cdot \vec{b}_\perp}\left[{\cal M}_{\rm fig.\ref{softpole}(d)}-{\cal M}_{\rm fig.\ref{softpole}(d)}^{\rm mirror}\right] \ .
\end{align}
Here, the mirror diagram represents the amplitude with gluon attachment to the right side of the cut-line in Fig.~\ref{softpole}(d). We can further apply $\vec k_\perp=\vec k_{q2\perp}-\vec k_{1\perp}^L$ for Fig.~\ref{softpole}(d) and $\vec k_\perp=\vec k_{q1\perp}-\vec k_{1\perp}^R$ for its mirror graph. Note that $k_1^L$ and $k_1^R$ are different because they have different $k_{q1\perp}$ and $k_{q2\perp}$ dependences. We find that the Sivers function in $b_\perp$-space is proportional to
\begin{align}
&\tilde{f}_{1T}^\perp(x,b_\perp,\mu,\zeta_z)|_{\rm fig.\ref{softpole}(d)}\propto \nonumber \\ &e^{i\vec{k}_{q2\perp}\cdot \vec{b}_\perp}\int\frac{d^4k_1^L}{(2\pi)^2}e^{i\vec{k}_{1\perp}^L\cdot \vec{b}_\perp}\frac{1}{(k_z^L)^2}{\cal R}(k_{1z}^L)\delta((k_1^L)^2)\nonumber\\
&-e^{i\vec{k}_{q1\perp}\cdot \vec{b}_\perp}\int\frac{d^4k_1^R}{(2\pi)^2}e^{i\vec{k}_{1\perp}^R\cdot \vec{b}_\perp}\frac{1}{(k_z^R)^2}{\cal R}(k_{1z}^R)\delta((k_1^R)^2)\ ,
\end{align}
where ${\cal R}(k_z)=\left(1-e^{ ik_zL}\right)\left(1-e^{ -ik_zL}\right)$. Notice that although $k_1^L$ and $k_1^R$ are not identical, their contribution to the above equation is the same. So, we can combine the above two terms and obtain,
\begin{align}
&\tilde{f}_{1T}^\perp(x,b_\perp,\mu,\zeta_z)|_{\rm fig.\ref{softpole}(d)}\propto
\left[e^{i\vec{k}_{q2\perp}\cdot \vec{b}_\perp}-e^{i\vec{k}_{q1\perp}\cdot \vec{b}_\perp}\right]\nonumber \\ &\times\int\frac{dk_{1z}}{k_{1z}^2} \frac{d^2k_{1\perp}}{(2\pi)^2} e^{i\vec{k}_{1\perp}\cdot \vec{b}_\perp}\frac{1}{\sqrt{k_{1z}^2+k_{1\perp}^2}}{\cal R}(k_{1z}) \ .
\end{align}
It is interesting to observe that the first factor is just the leading order expression and the second factor represents the amplitude without gluon attachments and is a diagram similar to that for the unpolarized quark distribution at one-loop order. Applying the result from Ref.~\cite{Ji:2018hvs}, we have
\begin{align}
&\tilde{f}_{1T}^\perp(x,b_\perp,\mu,\zeta_z)|_{\rm fig.\ref{softpole}(d)}\nonumber \\ &=-\frac{ib_\perp^\alpha}{2}\frac{\alpha_s}{2\pi}\frac{1}{2N_c}T_F(x,x) \int\frac{dk_z}{k_z^2}\frac{d^2k_\perp}{(2\pi)^2}e^{i\vec{k}_\perp\cdot \vec{b}_\perp}\frac{{\cal R}(k_z)}{\sqrt{k_z^2+k_\perp^2}} \nonumber\\
&=-\frac{ib_\perp^\alpha}{2}\frac{\alpha_s}{2\pi}\frac{1}{2N_c}T_F(x,x) 2{\cal K}(\xi_b) \ ,
\end{align}
where $\xi_b=L/|{\vec b}_\perp|$ and the function ${\cal K}$ is defined as~\cite{Ji:2018hvs},
\begin{equation}
{\cal K}(\xi_b)=2\xi_b \tan^{-1}\xi_b-\ln(1+\xi_b^2) \ .
\end{equation}
At large $\xi_b$ the above ${\cal K}(\xi_b)$ becomes $\pi\xi_b-2\ln\xi_b$, while
at small $\xi_b$ it behaves as $\xi_b^2$.

\begin{figure}[tbp]
\begin{center}
\includegraphics[width=8cm]{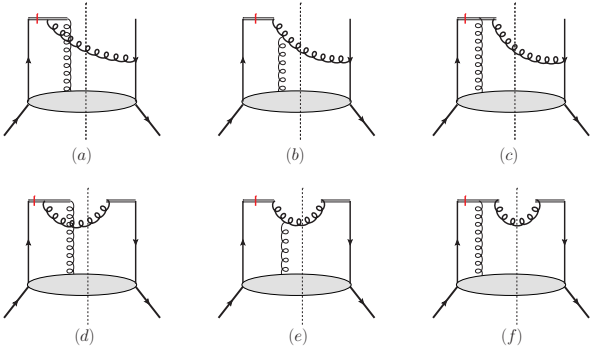}
\end{center}
\caption[*]{Hard pole contributions at one-loop order for quasi-Sivers function.}
\label{hardpole}
\end{figure}

Now, let us move to hard-gluon-pole contributions, for which the diagrams are shown in Fig.~\ref{hardpole}. First, we notice that the phase contribution comes from the gauge link propagator,
\begin{equation}
\frac{1}{n_z\cdot (k_{q1}-k)}=\frac{1}{n_z\cdot P}\frac{1}{x_{q1}-x} \ .
\end{equation}
The pole contribution leads to $\delta(x_{q1}-x)$ which means that $x_g=(1-\xi_x)x_q$. Because the pole is situated at  $x_g\neq 0$ this is a hard-gluon pole contribution. Similar as for the standard Sivers function, these diagrams
do not produce derivative terms. Again, we apply the collinear expansion of the incoming parton transverse momenta and for convenience we have chosen the physical polarization for the radiated gluon in these diagrams. The total contribution from Fig.~\ref{hardpole} can be separated into two terms: one contains the pinch-pole singularity and one is free of pinch-pole singularity.

We can derive the pinch-pole term following the above procedure, giving the contribution from the soft gluon pole diagrams. Again, we have to use a finite length to regulate the divergence and the result in $b_\perp$-space is the same as above but with a different color factor,
\begin{align}
&\tilde{f}_{1T}^\perp(x,b_\perp,\mu,\zeta_z)|_{\rm fig.\ref{hardpole}}^{\rm pinch~p.}\nonumber\\ &=\frac{ib_\perp^\alpha}{2}\frac{\alpha_s}{2\pi}\frac{C_A}{2}T_F(x,x)\int\frac{dk_z}{k_z^2}\frac{d^2k_\perp}{(2\pi)^2}e^{i\vec{k}_\perp\cdot \vec{b}_\perp}\frac{{\cal R}(k_z)}{\sqrt{k_z^2+k_\perp^2}} \nonumber\\
&=\frac{ib_\perp^\alpha}{2}\frac{\alpha_s}{2\pi}\frac{C_A}{2}T_F(x,x) 2{\cal K}(\xi_b) \ .
\end{align}

After subtracting the pinch-pole contribution, we derive the rest of the hard gluon pole contribution from Fig.~\ref{hardpole},
\begin{align}
&\tilde f_{1T}^\perp(x,k_\perp,\mu,\zeta_z)|_{\rm fig.\ref{hardpole}}^{\rm no-pinch~ p.}\nonumber \\ &=-\frac{M_p}{(k_\perp^2)^2}\frac{\alpha_s}{2\pi^2}\frac{C_A}{2}\int\frac{dx_q}{x_q}T_F(x,x_q)\frac{(1-\xi)^2(1+\xi_x)}{(1-\xi_x)^2\sqrt{1+r_\perp^2}}\ ,
\end{align}
where we have neglected power corrections in the TMD limit, and we have applied the symmetry property of the Qiu-Sterman matrix element $T_F(x,x_q)=T_F(x_q,x)$ to simplify the final result. In the TMD limit, the expression above can be further simplified to
\begin{align}
&\tilde f_{1T}^\perp(x,k_\perp,\mu,\zeta_z)|_{\rm fig.\ref{hardpole}}^{\rm no-pinch~ p.}\nonumber \\ &=-\frac{M_p}{(k_\perp^2)^2}\frac{\alpha_s}{2\pi^2}\frac{C_A}{2}\int\frac{dx_q}{x_q}T_F(x,x_q)\nonumber \\ &\times\left[\frac{1+\xi_x}{(1-\xi_x)_+}+\delta(1-\xi_x)\ln\frac{\zeta_z}{k_\perp^2}\right]\ .
\end{align}

Similar to the case for the standard quark Sivers function calculated in Ref.~\cite{Ji:2006vf}. We observe the following: There are cancellations between the hard gluon pole contributions and the soft gluon pole contributions. In particular, the hard-gluon-pole is proportional to the color factor $C_A/2$ while the soft-gluon-pole to $1/2N_c$. Their cancellation leads to the final result proportional to $C_F$ for the end-point contribution and the pinch-pole contributions. This is consistent with the soft gluon radiation contribution and the soft factor subtraction.

Since the soft factor and the subtraction is spin-independent, their contributions will be same as those calculated in Ref.~\cite{Ji:2018hvs}. Combining all these terms, we obtain the final result for the quark Sivers function at one-loop order in LaMET,
\begin{align}
&\tilde{f}_{1T}^{\perp\alpha(1)}(x,{\vec b}_\perp,\mu,\zeta_z)\nonumber \\ &=\frac{ib_\perp^\alpha}{2}\frac{\alpha_s}{2\pi}\left\{\left(-\frac{1}{\epsilon}
-L_b\right){\cal P}_{qg/qg}^T\otimes T_F(x,x)\right.\nonumber\\
&\left.+\int\frac{dx_q}{x_q}T_F(x_q,x_q)\left[-\frac{1}{2N_c}(1-\xi_x)+
\delta(1-\xi_x)C_F\tilde s^{(1)}\right]\right\}\ ,
 \label{oneloop}
\end{align}
where ${\cal P}_{qg/qg}^T\otimes T_F(x,x)$ has been defined in Eq.~(\ref{ptt}) and $\tilde s^{(1)}$ is the same as for the unpolarized case of Eq.~(\ref{h1}). Similar to the unpolarized case, after renormalization, we can write the quasi Sivers function in terms of the collinear twist-three Qiu-Sterman matrix element,
\begin{align}
&\tilde{f}_{1T}^{\perp\alpha}(x,{\vec b}_\perp,\mu_b,\zeta_z)\nonumber \\ &=\frac{ib_\perp^\alpha}{2}\int\frac{dx_q}{x_q}T_F(x_q,x_q,\mu_b)\left\{\delta(1-\xi_x)\right.\nonumber\\
&\left.+\frac{\alpha_s}{2\pi}\left[-\frac{1}{2N_c}(1-\xi_x)+\delta(1-\xi_x)C_F\tilde s^{(1)}\right]\right\} \ ,\label{siversoneloop}
\end{align}
at one-loop order.

A couple of comments are in order before we close this section. First, the above result is obtained in the scheme of Ref.~\cite{Ji:2019ewn} in which $\sqrt{Z_E}$ was adopted to subtract out the pinch-pole singularity. The same strategy has already been adopted in Ref.~\cite{Ji:2018hvs} in which it is regarded as a soft factor subtraction.

Second, for the Sivers contribution, we focused our calculations for the soft-gluon and hard-gluon pole contributions from the Qiu-Sterman matrix element $T_F(x_1,x_2)$. Other contributions from the soft-fermion pole and those from $\tilde G_F(x_1,x_2)$ can be included as well. They contribute to both the evolution kernel ${\cal P}_{qg/qg}^T$ and the finite term of Eq.~(\ref{siversoneloop}), see, for example, the recent study for the light-cone Sivers function in Ref.~\cite{Scimemi:2019gge}.

\section{Single Transverse-Spin Asymmetry}

In this section, we discuss applications of the results obtained in previous sections. In particular, we consider the single spin asymmetry at large and small $b_\perp$. We also comment on previous lattice calculations
for moments of the relevant TMDPDFs.

\subsection{Single Spin Asymmetry}
One of the most important physical application of spin-dependent TMDPDFs are single-spin asymmetries defined by the ratio of physical cross sections. For example, in the Drell-Yan lepton pair production, we define, 
\begin{align}
   A_{DY}
   = \frac{\frac{d^4\sigma_{+S_\perp}}{d^2Q_\perp dx_Adx_B}-\frac{d^4\sigma_{-S_\perp}}{d^2Q_\perp dx_Adx_B}}{\frac{d^4\sigma_{+S_\perp}}{d^2Q_\perp dx_Adx_B}+\frac{d^4\sigma_{-S_\perp}}{d^2Q_\perp dx_Adx_B}} \ ,
\end{align}
where $\frac{d^4\sigma_{\pm S_\perp}}{d^2Q_\perp dx_Adx_B}$ are the differential cross sections with the transverse spin for the polarized target being $\pm S_\perp$, $x_{A,B}$ denote the momentum fractions of the incoming hadrons carried by the quark and antiquark, $Q^2$ and $Q_\perp$ are the invariant mass and transverse momentum for the lepton pair.
The factorization formula~\cite{Collins:2011zzd} for the Drell-Yan or SIDIS process in terms of the physical TMDPDFs reads
\begin{align}
    &\frac{d^4\sigma}{dQ_\perp^2dx_Adx_B}=\hat \sigma(\frac{Q^2}{\mu^2}) \times \nonumber \\ &\int d^2b_\perp e^{i\vec{b}_\perp \cdot \vec{Q}_\perp}q(x_A,b_\perp,S_\perp,\mu,\zeta_A)q(x_B,b_\perp,\mu,\zeta_B)
\end{align}
where 
$x_A$ is the momentum fraction for the quark parton coming out of the polarized target and $x_B$ is for the unpolarized one. $\hat \sigma(\frac{Q^2}{\mu^2})$ is the hard-cross section with $Q^2=2x_Ax_BP^+P^-$ where $P^{\pm}$ are the largest light-front components of the hadron momenta. Given the factorization formula and utilizing a decomposition for the TMDPDF similar to Eq.~(\ref{eq:decompo}), one found that the single-spin asymmetry for the Drell-Yan process can be written in terms of an unpolarized TMDPDF and Sivers function:
\begin{align}
    &A_{DY}=\nonumber \\ &\frac{\int d^2b_\perp e^{i\vec{Q}_\perp\cdot \vec{b}_\perp}\epsilon^{\beta\alpha}S_\perp^\beta f_{1T}^{\perp\alpha}(x_A,b_\perp,\mu,\zeta_A)q(x_B,b_\perp,\mu,\zeta_B)}{\int d^2b_\perp e^{i\vec{Q}_\perp\cdot \vec{b}_\perp}q(x_A,b_\perp,\mu,\zeta_A)q(x_B,b_\perp,\mu,\zeta_B)} 
\end{align}
where the hard cross-section induced by unpolarized quark partons cancels between numerator and denominator. By using the matching relation Eq.~(\ref{eq:factorization}) and choosing $\zeta_{zA}=\zeta_A=2x_A^2(P^+)^2$, $\zeta_{zB}=\zeta_B=4x_B^2(P^-)^2$, one has
\begin{align} \label{eq:singleas}
    &A_{DY}=\frac{D}{S} \ ,\\
    &D=\int d^2b_\perp e^{i\vec{Q}_\perp\cdot \vec{b}_\perp}\epsilon^{\beta\alpha}S_\perp^\beta \tilde f_{1T}^{\perp\alpha}(x_A,b_\perp,\mu,\zeta_{zA})\nonumber \\ &\times \tilde q(x_B,b_\perp,\mu,\zeta_{zB})S_r(b_\perp,\mu) \ , \\ 
    &S=\int d^2b_\perp e^{i\vec{Q}_\perp\cdot \vec{b}_\perp}\tilde q(x_A,b_\perp,\mu,\zeta_{zA})\tilde q(x_B,b_\perp,\mu,\zeta_{zB})\nonumber \\ &\times  S_r(b_\perp,\mu)  \ .
\end{align}
Notice that the matching kernel all cancel, but the reduced soft function $S_r$ does not cancel between $D$ and $S$. Since $\tilde q$, $\tilde f_{1T}^{\perp\alpha}$ and $S_r$ can all be extracted from lattice calculations, Eq.~(\ref{eq:singleas}) allows to predict the physical observable single-spin asymmetry from lattice data. 

Because $A_{DY}$ is complicated due to the fact that the soft contribution fails to cancel, it is attractive to mimic $A_{DY}$ by using the following simplified version for the  single-spin asymmetry ratio in Fourier transform $b_\perp$-space:
\begin{align}\label{eq:singlespin}
&{\cal R}_{S_\perp}(x,b_\perp)\nonumber \\ &=\frac{q(x,b_\perp,S_\perp,\mu,\zeta)-q(x,b_\perp,-S_\perp,\mu,\zeta)}{q(x,b_\perp,S_\perp,\mu,\zeta)+ q(x,b_\perp,-S_\perp,\mu,\zeta)} \nonumber \\ &=\epsilon^{\beta\alpha}S_\perp^\beta\frac{f_{1T}^{\perp\alpha}(x,b_\perp,\mu,\zeta)}{q(x,b_\perp,\mu,\zeta)}\ .
\end{align}
Instead of transforming to momentum space, in ${\cal R}_{S_\perp}$ 
we directly compare the asymmetry point by point in $b_\perp$ space. We emphasize that this asymmetry is not a physical observable, but a ratio between the quark-Sivers function and the unpolarized quark distribution. 
An important feature is that the $\mu$, $\zeta$ dependencies cancel between numerator and denominator, thus the ${\cal R}_{S_\perp}$ 
is independent of the renormalization scale $\mu$ and the rapidity scale $\zeta$. This contribution is proportional to $\vec{S}_\perp\times \vec{b}_\perp$ and the coefficient defines the size of the single spin asymmetry in the quark distribution. The individual TMDPDFs will depend on the rapidity renormalization scheme. However, the ratio between the quark Sivers function and the unpolarized quark distribution does not depend on the scheme. In particular, the scheme dependent soft functions cancel in the ratio of Eq.~(\ref{eq:singlespin}).

With the relation between the quasi-TMDPDF and physical TMDPDF in Eq.~(\ref{eq:factorization}), we will be able to study the single spin asymmetry in the quark distribution. Using the fact that the matching kernel $H$ is independent of the spin-structure and is the same for both the Sivers function and the un-polarized quark TMDPDF, we found that by taking the ratio, the soft factor and matching kernel dependencies cancel:
\begin{align}\label{eq:singlespinquasi}
\widetilde{{\cal R}}_{S_\perp} 
&=\epsilon^{\beta\alpha}S_\perp^\beta\frac{\tilde f_{1T}^{\perp\alpha}(x,b_\perp,\mu,\zeta_z)}{\tilde q(x,b_\perp,\mu,\zeta_z)}\nonumber \\
&=\epsilon^{\beta\alpha}S_\perp^\beta\frac{f_{1T}^{\perp\alpha}(x,b_\perp,\mu,\zeta)}{q(x,b_\perp,\mu,\zeta)}\equiv {\cal R}_{S_\perp} 
(x,b_\perp) \ .
\end{align}
Therefore, the single-spin asymmetry ratio extracted from the quasi-Sivers function and the Sivers function are the same. This is one of the major results of this paper.

At small $b_\perp$, the single spin asymmetry ratio in the quark distribution at one-loop order can be extracted from the perturbative results of the Sivers function provided in the previous sections
\begin{align}
&{\cal R}_{S_\perp} 
(x,b_\perp)|_{b_\perp\ll \frac{1}{\Lambda_{QCD}}}=\frac{i|b_\perp|\sin(\phi_b)}{2}\nonumber \\ &
\times \frac{T_F(x,x,\mu_b)-\frac{\alpha_s}{4\pi N_c} (1-x/x_q)\otimes T_F(x_q,x_q,\mu_b)}{f_q(x,\mu_b)+\frac{\alpha_sC_F}{2\pi} (1-x/x_q)\otimes f_q(x_q,\mu_b)} \ ,\label{tfan}
\end{align}
where $\phi_b$ is the azimuthal angle between $\vec{b}_\perp$ and $\vec{S}_\perp$ and we have performed the renormalization of the quark distribution at $\mu=\mu_b$. As one can see, the soft contribution in $s^{(1)}$ cancels. 
The explicit derivations in the previous sections have confirmed that the single spin asymmetry extracted from quasi-Sivers function are the same for small $b_\perp$.

 For the quark Sivers asymmetry, the $\alpha_s$ correction is very small. For example, numerically, the $\alpha_s$ corrections for the numerator and denominator are less than $1\%$ in most of kinematics of the valence quark distributions at $x\sim 0.2$. Therefore, we can safely neglect these corrections and interpret the asymmetry as the ratio between the Qiu-Sterman matrix element and the unpolarized quark distribution at the scale $\mu_b$.

\subsection{Asymmetry at Large $b_\perp$}

On the other hand, the spin asymmetry ratio at large-$b_\perp$ is determined by non-perturbative TMDPDFs, for which lattice calculations in terms of the formalism presented in Sec. II will be very important. 

In previous phenomenology studies, a Gaussian distribution in the transverse momentum space has been assumed for both the unpolarized and Sivers quark distributions, e.g., 
\begin{eqnarray}
q(x,k_\perp,\mu=\zeta=\mu_0)&\propto & e^{-\frac{k_\perp^2}{Q_0^2}}\ , \\
f_{1T}^\perp(x,k_\perp,\mu=\zeta=\mu_0)&\propto & e^{-\frac{k_\perp^2}{Q_s^2}}\ , 
\end{eqnarray}
where $Q_0$ and $Q_s$ are parameters for the Gaussian distributions. Because the asymmetry ratio has to decrease at large transverse momentum, the Gaussian width for the quark Sivers function is smaller than that for the unpolarized quark distribution, i.e., $Q_s<Q_0$. If we translate this into $b_\perp$-space, it will generate a significantly increasing function for ${\cal R}_{S_\perp} 
(x,b_\perp)$ at large $b_\perp$,
\begin{equation}
    {\cal R}_{S_\perp} 
    (b_\perp)|_{\rm model}\propto b_\perp e^{\frac{(Q_0^2-Q_s^2)b_\perp^2}{4}} \ .
\end{equation}
For example, the parameterization in Ref.~\cite{Sun:2013hua} predicts a factor of $25$ increase from $b_\perp=0.2~\rm fm$ to $b_\perp=1~\rm fm$. Similar predictions exist for other parameterizations, see, some recent global analyses~\cite{Cammarota:2020qcw,Bacchetta:2020gko}. It will be crucially important to check this in lattice simulations.

\subsection{Relation to Previous Lattice Simulations}

In Refs.~\cite{Musch:2010ka,Musch:2011er,Yoon:2017qzo}, lattice computations for certain matrix 
elements in a hadron state have been carried out. These matrix elements are defined through TMDPDF-like bi-local operators, which are separated by transverse distance $b_\perp$ perpendicular to the hadron's momentum direction. 
It is easy to see that the matrix elements calculated there do not correspond exactly to the moments of the quasi-TMDPDF distribution. Therefore, it is hard to interpret them although interesting results were obtained.

Our results, obtained by using LaMET can help to improve these earlier results. For example, we can add the explicite $x$ dependence to the matrix elements calculated in Refs.~\cite{Musch:2010ka,Musch:2011er,Yoon:2017qzo}. This can help to resolve the difference between, e.g., 
\begin{align}
&\frac{\int dx x^n \tilde f^{\perp \alpha}_{1T}(x,b_\perp,\mu,\zeta_z=4x^2(P^z)^2)}{\int dx x^n \tilde q(x,b_\perp,\mu,\zeta_z=4x^2(P^z)^2)}\nonumber \\ \ne &\frac{\int dx x^n f^{\perp \alpha}_{1T}(x,b_\perp,\mu,\zeta)}{\int dx x^n q(x,b_\perp,\mu,\zeta)} \ .
\end{align}
The above point has also been observed in Ref.~\cite{Ebert:2020gxr}. 
It will be useful to have lattice simulations in the LaMET framework to constrain the quark Sivers functions and compare to phenomenological studies~\cite{Sun:2013hua,Cammarota:2020qcw,Bacchetta:2020gko}.

\section{Conclusion}

In summary, we have investigated the quark Sivers function in LaMET. A number of important features have been found for these distribution functions. In our derivation, we adopted the definition of quasi-TMDPDF in Refs.~\cite{Ji:2018hvs,Ji:2019ewn}. We have shown that the quasi-Sivers function can be matched to the physical Sivers function using Eq.~(\ref{eq:factorization}). The matching kernel and the reduced soft function are the same as that of the unpolarized case. As a result, in the single spin asymmetry, the soft function and hard kernels cancel between the quark Sivers function and the unpolarized quark distribution and we can extract the physical single spin asymmetry knowing the  lattice calculable quasi-TMDPDFs.

As a byproduct, Eq.~(\ref{tfan}) provides another useful method to compute twist-three quark-gluon-quark correlation function, in particular, for those directly connected to the leading order TMDPDFs. 
We notice a recent study of twist-three parton distribution $g_T(x)$ in LaMET framework~\cite{Bhattacharya:2020xlt}. These studies demonstrate the powerful reach of the LaMET formalism and we hope that more studies of this type will become available in the future.

The methods discussed in this paper can be extended in various directions. 
An immediate extension is the analysis of all other $k_\perp$-odd quark TMDPDFs. The large transverse momentum dependence for these distributions has been derived in Ref.~\cite{Zhou:2009jm}. In order to study them in LaMET, we need to translate these results into the LaMET formalism following the procedure of the current paper for the quark Sivers function. We plan to study this in a future publication. Together with a recent paper on $k_\perp$-even spin dependent quark TMDPDFs~\cite{Ebert:2020gxr}, this will complete all leading quark TMDPDFs in LaMET.

In addition, the method developed in this paper shpold be applied to other related parton distribution functions, especially for those relevant for the quantum phase space Wigner distributions. These distribution functions contain in principle the complete information needed for nucleon tomography and allow to unveil the origin of the parton orbital angular momentum in nucleons. We expect more developments on this subject soon.

\vspace{2em}
We thank Jianhui Zhang and Yong Zhao for discussion and suggestions. This material is based upon work partially supported by the U.S. Department of Energy, Office of Science, Office of Nuclear Physics, under contract number DE-AC02-05CH11231 and DE-SC0020682, and Center for Nuclear Femtography, Southeastern Universities Research Associations in Washington DC, and within the framework of the TMD Topical Collaboration. AS acknowledges support by SFB/TRR-55.

\bibliographystyle{apsrev4-1}
\bibliography{bibliography}

\end{document}